\documentstyle[epsfig]{article}
\def\be{\begin{equation}}
\def\ee{\end{equation}}
\def\bea{\begin{eqnarray}}
\def\eea{\end{eqnarray}}
\def\ben{\begin{eqnarray}}
\def\enn{\end{eqnarray}}

\def\dst{\displaystyle\phantom{|}}
\def\l({\left(}
\def\r){\right)}

\def\bk{{\bf k}}
\def\bK{{\bf K}}

\begin{document}
\begin{center}
{\LARGE\bf
	Partial coherence in the core/halo picture \\[1ex]
	of Bose-Einstein $n$-particle correlations }\\[2ex]	
\end{center}
\medskip
\begin{center}
		{\large \sc
	T. Cs\"org\H o$^1$, B. L\"orstad$^2$, \\[1ex]
		J. Schmidt-S\o rensen$^2$ and A. Ster$^3$}\\[1.5ex]
\end{center}
\medskip
\begin{center}
{\large\it $^1$  MTA KFKI RMKI, H - 1525 Budapest 114, POB 49, Hungary \\
     $^2$  Physics Department, Lund University, \\
	S - 221 00 Lund, POB 118, Sweden\\
     $^3$  MTA KFKI MFA, H - 1525 Budapest 114, POB 49, Hungary }\\[4ex]
\end{center}
\medskip
\bigskip
\begin{center}
	{\bf Abstract}\\
\begin{minipage}[t]{10cm}
We study the influence of a possible coherent component in the boson source 
on the two-, three- and
$n$-particle correlation functions in a generalized core/halo type of
boson-emitting source. In particular, a simple formula 
is presented for the strengh of the $n$-particle correlation
functions for such systems.  Graph rules are obtained 
to evaluate the correlation functions of arbitrary high order. 
The importance of experimental determination of the 4-th and 5-th
order Bose-Einstein correlation function is emphasized.
\end{minipage}
\end{center}
\bigskip
\bigskip

\section{Introduction}
	Intensity correlations were discovered first in
	astrophysics by R. Hanbury Brown and  R. Q. Twiss~\cite{hbt},
	who invented this method to	determine the angular diameter
	of main sequence stars (HBT effect). In particle physics, intensity
	correlations of pions were observed by Goldhaber, Goldhaber,
	Lee and Pais (GGLP effect)~\cite{gglp}.
	Bose-Einstein correlations are intensity
	correlations among detected bosons,  that are studied 
	mainly with the purpose of reconstructing the 
	space-time picture of particle production.
	The analysis of higher-order Bose-Einstein 
	correlation functions became a focal point of current
	research interest. 

	In particle physics, significant three or higher order 
	Bose-Einstein correlations have been extracted from the data
	sampled by the AFS~\cite{afs_n1},
	the NA22~\cite{na22_n1,na22_n2,na22_n3} 
	and the UA1-collaborations ~\cite{ua1_n1}. 
	These data were used to test the possible existence of
	a coherent source in multi-particle physics and to compare
	the correlation functions to the 
	strength of these correlations predicted from
	the quantum optical QO formalism~\cite{glauber,qo,qow}. 
	As the precision of the measurements improved, the 
	QO predictions with higher order correlations
	were found to be marginally consistent with the 
	data on 3 and 4 order Bose-Einstein correlation
	functions (BECF-s)~\cite{na22_ng} in $(\pi^+/K^+) + p$ reactions
	at CERN SPS. Recently, this basic QO formalism  was shown 
	be insufficient to simultaneously describe the high precision
	UA1 data on two- and three-particle Bose-Einstein correlations
	~\cite{ua1_ng}.

	In high energy heavy ion physics, 
	the first experimental determination of the three-particle
	correlation function has just been 
	reported by the NA44 collaboration
	~\cite{na44-qm97,bengt-cf98,janus-cris98,na44-3pi}.
	NA44 reports that the  genuine three-particle correlation is
	quite suppressed in the studied reaction, $S+Pb$ collisions.
	By the genuine three-particle correlation is meant the part
	of the three-particle correlation that is not due to included
	combinations of two-particle correlations. This suppression can
	be expressed as a phase factor,  $\cos(\phi)$, of the three-particle
	correlation function in the case of totally incoherent particle
	production. In that case this phase factor is related to an
	asymmetry of the particle source, not possible to extract
	from two-particle correlations. Theoretical estimates
	of this asymmetry effect on the phase factor show very small
	departures from $\cos(\phi) \approx 1.$~\cite{3pi-nbi,henning,uli_zhang}.
	The large departure from  $\cos(\phi) =1$ found by the NA44 collaboration,
	$\cos(\phi) =0.2 \pm 0.2$~\cite{na44-3pi}, ought to be due to some other mechanism.
	We will discuss the possibility of a partial coherent source in this Letter.
	A possible existence of
	such an extra phase in the three and higher order correlation
	functions was noted already e.g. in papers by the
	NA22 collaboration~\cite{na22_ng}, but no experimental
	evidence has been put forward for a $\cos(\phi) \ne 1$ value 
	in particle physics.

	From the theoretical side, Cramer and Kadija predicted up to
	order 6 the strength of Bose-Einstein correlations for sources with 
	partially coherent and incoherent components that included 
	also a possible contamination by mis-indentified, non-interfering
	particles~\cite{cramer-kadija}. Their formulas were 
	obtained in the quantum-optical formalism.
	Recently, Suzuki and collaborators calculated higher order
	exclusive Bose-Einstein correlations from 
	the generating functional approach to
	quantum-optical formalism~\cite{suzuki} for the case that
	the source has $M$ incoherent and  one coherent component.

	Recently, multi-particle symmetrizations up to arbitrary 
	high order were evaluated exactly by Zhang~\cite{zhang-npi}
	for the special case of a pion-laser model proposed by Pratt
	in ref.~\cite{plaser}.
	Surprizingly, the structure of the $n$-particle inclusive
	correlation functions in terms of the Fourier-transformed
	inclusive emission function 
	was found to be the same as the structure of the $n$-particle 
	exclusive correlation functions in terms of the 
	single-particle exclusive emission function~\cite{zhang-npi}.
	However, this result is valid only in case 
	when Bose-Einstein condensation, hence the development of
	 partial coherence, is not yet reached~\cite{zjcst-rev}.

	A simple recurrence relation was obtained 
	for the strength of the higher order correlation functions
	of core/halo type systems~\cite{nhalo}. 
	Such systems are boson emitting
	sources where some particles come from the a incoherent center of the
	particle emission, that is assumed to be resolvable 
	by the Bose-Einstein microscope. The rest of the particles
	is assumed to come from the halo region, that corresponds
	to large length-scales not resolvable by intensity interferometry
	~\cite{chalo,marburg-halo}.
	In Letter ~\cite{nhalo}, a prediction was made for the strength
	of third order and arbitrary order BECF   assuming that the 
	core has no coherent component. 
	
	The purpose of the present Letter is to investigate the 
	effect of a partially coherent component in the core of
	particle emission. We present a generalization of
	the earlier recurrence relations in ref.~\cite{nhalo},
	the new expressions also yield an easy way to calculate formula for the
	strength of the $n$-th order correlation function with a
	partially coherent and a halo component and we apply
	these expressions to the NA44 data on $S+Pb$ collisions.

\section{Basic definitions}

	The central assumption of the core/halo model is 
	that the  reduction of the intercept parameter of the $n$-particle
	BECF-s
	is due only to the presence of the long-lived resonances~\cite{nhalo}.
	This assumption was motivated by the success of fully incoherent
	event	generators like RQMD or VENUS in the description
	of two-particle BECF-s.

	The emission function of the whole source
	can be written as a sum of a contribution
	from the core and from the halo, where
	halo stands for the decay products of the (unresolvable)
	long-lived resonances. The core is indexed with (c) ,
	the halo is by (h). 
\be
	S(x,k)  =  S_c(x,k) + S_h(x,k)
\ee

	In earlier studies of the core/halo model it was assumed
	that $S_c(x,k)$ describes a fully incoherent (thermal)
	source. Now we assume, that some fraction of the core
	emits bosons in a coherent manner, e.g., due to 
	emerging formation of pion lasers or
	Bose-Einstein condensates of pions or production of disoriented
	chiral condensates or ..., so we define

\be
	S_c(x,\bk)  =  S_c^p(x,\bk) + S_c^i(x,\bk)
\ee
	where the upper index $p$ stands for  coherent
	component (p as partial), upper index $i$ stands for incoherent
	component of the source.

	The invariant spectrum is given by 
\be
	N(\bk) = \int d^4x S(x,\bk) = N_c(\bk) + N_h(\bk)
\ee
	and the core contribution is a sum of the coherent
	and incoherent components:
\be
	N_c(\bk) = \int d^4x S_c(x,\bk) = N_c^p(\bk) + N_c^i(\bk)
\ee
	One can introduce the momentum dependent core fractions $f_c(\bk)$
	and partially coherent core fractions $p_c(\bk)$ as
\bea
	f_c(\bk) & = & N_c(\bk)/N(\bk) \label{e:fck} \\
	p_c(\bk) & = & N_c^p(\bk) / N_c(\bk)
\eea
	The halo and the incoherent fractions $f_h, f_i$ are 
\bea
	f_h(\bk) & = & N_h(\bk)/N(\bk) = 1 - f_c(\bk) \\
	f_i(\bk) & = & N_c^i(\bk) / N_c(\bk) = 1 - p_c(\bk)
	\label{e:fik}
\eea

	Note that our definition of the momentum dependent,
	partially coherent core fraction, $p_c(\bk)$ 
	should be clearly distinguished
	from the chaociticy $p$ of Weiner~\cite{weiner},
	defined as $p = \langle n_{chao}\rangle / \langle n_{tot} \rangle$,
	the ratio of the mean number of particles from the chaotic source
	to the mean total multiplicity. If we neglect the momentum dependence
	of $f_c(\bk)$ and $p_c(\bk)$, the core fraction and the partially
	coherent core fraction, formally one obtains $p = 1 - p_c f_c$.
	However, we distinquish the resolvable intercept $\lambda_*$ from
	the exact intercept $\lambda_{xct}$, in contrast to ref.~\cite{weiner}.
	For example, in case of two-particle correlations,
	$\lambda_{*,2} =   f_c^2 [ (1 - p_c)^2 + 2 p_c (1 - p_c)]$,
	while $\lambda_{xct,2} = \lambda_{*,2} + (1 - f_c)^2 +
	2 f_c (1 - f_c)$ in our case, while in case of the quantum
	optical formalism without long lived resonances,
	$\lambda_2^{QO}  =   2 p (1 - p) + p^2 = 1 - (1 - p)^2$.

\section{ The strength of the $n$-particle correlations, $\lambda_{*,n}$}
We define the $n$-particle correlation function as
\bea
	C_n(1,2,..., n) & = &C_n(\bk_1,\bk_2,...,\bk_n) =
	\frac{N_n(\bk_1,\bk_2,...,\bk_n) }{N_1(\bk_1) N_1(\bk_2) ... N_1(\bk_n)} \\
	\null & = &
	\frac{N_n(1,2,...,n) }{N_1(1) N_1(2) ... N_1(n)}
\eea
	where a symbolic notation for $\bk_i$ is introduced,
	only the index of $\bk$ is written out in the argument.
	In the forthcoming, we shall apply this notation consistently
	for the arguments of various functions of the momenta,
	i.e., $f(\bk_i,\bk_j, ... , \bk_m)$ is symbolically
	denoted by $f(i,j, ... ,m)$. 

	We find that the intercept of the $n$-particle 
	correlation function ( extrapolated from
	finite relative momenta to  zero
	relative momentum) is given by 
	the following formula,
	
\be
	C_n(k_i=k_j,\forall i,j) = 1 + \lambda_{*,n}= 1 + \sum_{j=2}^n 
		{\left( \null^{\dst n}_{\dst  j}\right)}
			\alpha_j f_c^j 
	\left[ (1-p_c)^j + j p_c (1-p_c)^{j-1} \right] ,
	\label{e:lamnfp_sim}
\ee
	where $\alpha_j $ counts the number
	of fully mixing permutations of $j$ elements.  
	This can be calculated from a simple recurrence,
	as obtained in ref.~\cite{nhalo}.	

	Note that the equations of ref.~\cite{nhalo,chalo}
	were given for the purely incoherent core, and they 
	are modified above for an additional coherent
	component in a straight-forward manner. In general,
	terms proportional to $f_c^j$ in the incoherent case
	shall pick up an additional factor 
	$ [ (1 - p_c)^j + j p_c (1 - p_c)^{j-1} ]$
	in case the core has a coherent component.
	This extra factor means that either all $j$ particles
	must come from the incoherent part of the core, 
	or one of them can come from the coherent, the remaining
	$j-1$ particles from the incoherent part.
	If two or more particles come from the coherent component
	of the core, the contribution to intensity correlations
	vanishes as the intensity correlator for two coherent particles
	is zero.

Let us indicate the number of permutations that completely mix  exactly
$j$ non-identical elements by $\alpha_j$. There are
exactly $\left( \null^{\dst n}_{\dst  j} \right)$ different
ways to choose $j$ different elements from among $n$ different elements.
Since all the $n!$ permutations can be written as a sum over the
fully mixing permutations, the counting rule yields
a recurrence relation for $\alpha_j$, ref. ~\cite{nhalo}:
\bea
	\alpha_n & = & n! - 1 - \sum_{j = 1}^{n - 1}
	{\left( \null^{\dst n}_{\dst  j}  \right)} \alpha_j.
	\label{e:alp}
\eea
The first few values of $\alpha_j$ are given as
\bea
	\alpha_1 & = & 0, \\ 
	\alpha_2 & = & 1, \\ 
	\alpha_3 & = & 2, \\ 
	\alpha_4 & = & 9, \\ 
	\alpha_5 & = & 44, \\
	\alpha_6 & = & 265.
\eea
We have the following explicit expressions for the first few intercept
parameters:
\bea 
	\lambda_{*,2} & = & f_c^2
		[ (1 - p_c)^2 + 2 p_c (1-p_c)]
		\label{e:l2} \\
	\lambda_{*,3} & = & 3 f_c^2 
		[ (1 - p_c)^2 + 2 p_c (1-p_c)]  \nonumber
	\\
	\null & \null & \qquad
		+ 2 f_c^3 
		[ (1 - p_c)^3 + 3 p_c (1-p_c)^2]
		\\
	\lambda_{*,4} & = & 6 f_c^2 
		[ (1 - p_c)^2 + 2 p_c (1-p_c)]  \nonumber
	\\
	\null & \null & \qquad
		+ 8 f_c^3
		[ (1 - p_c)^3 + 3 p_c (1-p_c)^2] \nonumber
	\\
	\null & \null & \qquad
		+ 9 f_c^4
		[ (1 - p_c)^4 + 4 p_c (1-p_c)^3]
		\\
	\lambda_{*,5} & = & 
		10 f_c^2 
		[ (1 - p_c)^2 + 2 p_c (1-p_c)] \nonumber
	\\
	\null & \null & \qquad
		+ 20 f_c^3
		[ (1 - p_c)^3 + 3 p_c (1-p_c)^2] \nonumber
	\\
	\null & \null & \qquad
		+ 45 f_c^4 
		[ (1 - p_c)^4 + 4 p_c (1-p_c)^3]  \nonumber
	\\
	\null & \null & \qquad
		+ 44 f_c^5 
		[ (1 - p_c)^5 + 5 p_c (1-p_c)^4]
	\label{e:lamfp)}
\eea
	In the above equations, the effective intercept parameters,
	the core fraction and the partially coherent fraction
	are evaluated at a  mean momentum ${\bf K}$, 
	$\lambda_{*,n} = \lambda_{*,n}({\bK})$,
	$f_c = f_c({\bK})$ and $p_c = p_c({\bK})$. 

\section{The $n$-body correlation function }
		
	Let us give the closed form for the full correlation
	function for arbitrary high order of correlation functions,
	generalizing the results of ref.~\cite{nhalo} for an
	additional partially coherent component in the source:

	Let $\rho^{(n)}$ 
	stand for those permutations of $(1,...,n)$
	that are mixing  {\it all} the  
	numbers from 1 to $n$ and let us indicate by
	$\rho_i$ the value which is  
	replaced by $i$ in a given permutation belonging
	to the set of permutations $\rho^{(n)}$.
	(Superscript indexes a set of
	permutations, subscript stands for a given value).
	Then we have $\rho_i \ne i$ for all values of $i = 1,..., n$.

	If the partial coherent component is 
	vanishing, the general expression for 	
	the $n$-particle inclusive correlation function
	$C_n(1,...,n)$  was given
	in ref~\cite{nhalo} as
\ben
	C_n(1,...,n) & = & 1 + \sum_{j = 2}^n~~
	\sum_{i_1, ..., i_j = 1}^{\null \,\,\, n \,\,\, _{\prime}}~~
	\sum_{\rho^{(j)}} \prod_{k=1}^j
	f_c(i_k) \tilde s_c(i_k,i_{\rho_k}).
	\label{e:fmix}
\enn
	Here $\sum'$ indicates that the 
	summation should be taken over  those set of
	values of the indices which do not contain any value more than once,
 	and the Fourier-transformed emission function of the core
is
\ben
\tilde s_c(i,j)
= {\tilde S_c(i,j) \over \tilde S_c(i,i)}
\enn
\ben
\tilde s_c(i,j)
= 
\tilde s_c^*(j,i) {\tilde S_c(j,j) \over \tilde S_c(i,i)}
\ne \tilde s^*_c(j,i),  \label{e:asym}
\enn 

In the above equations, the tilde denotes Fourier-transformation
over the relative momenta,
\ben
\tilde S_c(l,m) = \int d^4 x \exp[i (k_l - k_m) \cdot x ] 
S_c(x,{k_l+k_m \over 2})
\enn
and similar expressions hold for the coherent and the
incoherent components of the core.
	\footnote{Note that with this definition the normalized Fourier-transformed
emission  function becomes asymmetric to the exchange of the
arguments and complex conjugation:
 although the relationship
$\tilde S_c(i,j) = \tilde S_c^*(j,i)$ is satisfied.}

The expression in eq.~(\ref{e:fmix}) is valid not only 
for the case when  exactly $n$
bosons are in the system, full symmetrization is performed,
$C_n(1,2,...,n)$ stands for the $n$-particle exclusive
correlation function and
$\tilde S_c(i,j)$ stands for the Fourier-transformed
core emission function without
modifications due multi-particle symmetrization~\cite{zhang-npi,zjcst-rev}. 
In addition, eq.~(\ref{e:fmix}) is also valid
when the only source of correlations between pions is
	due to Bose-Einstein symmetrization,
the number of pions is randomly varying from event to event,
and $C_n(1,2, ... ,n)$ is interpreted as the $n$-particle
inclusive correlation function~\cite{zhang-npi,zjcst-rev},
and $\tilde S_c(i,j) $  includes all higher order symmetrization
effects.
However, eq.~(\ref{e:fmix}) is valid only if the core has no partially
coherent component. 
If a coherent component is present,
one can introduce the normalized {\underline i}ncoherent and {\underline p}artially
coherent core fractions as
\bea
\tilde s_c^i(j,k)
& = & {\tilde S^i_c(j,k) \over \tilde S^i_c(j,j)}\\
\tilde s_c^p(j,k)
& = & {\tilde S^p_c(j,k) \over \tilde S^p_c(j,j)}
\eea

	and we obtain
\ben
	C_n(1,...,n) & = & 1 + \sum_{j = 2}^n~~
	\sum_{m_1, ..., m_j = 1}^{\null \,\,\, n \,\,\, _{\prime}}~~
	\sum_{\rho^{(j)}} 
	\left\{
	\prod_{k=1}^{j}
	f_c(m_k) [1 - p_c(m_k)] \,\, \tilde s^i_c(m_k,m_{\rho_k})
	\right. \nonumber
	\\
	\null & \null & \null \hspace{-2cm} 
	\left. +
	 \sum_{l = 1}^j f_c(m_l) p_c(m_l)  \,\, \tilde s^p_c(m_l,m_{\rho_l}) \!
	\prod_{k=1, k \ne l}^j
	f_c(m_k) [1 - p_c(m_k)] \, \tilde s^i_c(m_k,m_{\rho_k})
	\right\}
	\label{e:fpmix}
\enn
	This expression contains phases in the Fourier-transformed,
	normalized source distributions. Actually, two (momentum dependent) 
	phases  are present: 
	one denoted by $\phi^i(\bk_m,\bk_n)$ in the Fourier-transformed
	normalized {\it incoherent} core emission function, 
	$\tilde s_c^i(\bk_m,\bk_n)$ and another independent
	phase denoted by $\phi^c(\bk_m,\bk_n)$ is present in the 
	the Fourier-transformed
	normalized {\it coherent} core emission function,
	$\tilde s_c^p(\bk_m,\bk_n)$. One can write
\bea
	\tilde s_c^i(\bk_m,\bk_n)
	& = & |\tilde s_c^i(\bk_m,\bk_n)| \exp[i \phi^i(\bk_m,\bk_n)],\\
	\tilde s_c^p(\bk_m,\bk_n)
	& = & |\tilde s_c^p(\bk_m,\bk_n)| \exp[i \phi^p(\bk_m,\bk_n)].
\eea
	The shape of both the coherent and the incoherent components
	is arbitrary in these equations, but should correspond to the space-time distribution
	of particle production.
	 If the variances of the core are finite, 
	the emission functions are usually
	 parameterized by Gaussians.
	If the core distributions have power-law like tails, 
 	like in case of the Lorentzian distribution~\cite{3d},  
	then the Fourier-transformed emission functions correspond
	to exponentials or to power-law structures~\cite{bialas}.
	For completeness, we list these possibilities below:
\bea
	|\tilde s_c^i(\bk_m,\bk_n)|^2
	& = & \exp( - R_{i}^2 Q_{mn}^2) \qquad \mbox{\rm or} \\ 
	|\tilde s_c^i(\bk_m,\bk_n)|^2
	& = & \exp( - R_{i} Q_{mn})  \qquad \mbox{\rm or} \\ 
	|\tilde s_c^i(\bk_m,\bk_n)|^2
	& = & a_i  (R_{i} Q_{mn})^{b_i}  \qquad \mbox{\rm etc ... , } \\ 
	|\tilde s_c^p(\bk_m,\bk_n)|^2
	& = & \exp( - R_{p}^2 Q_{mn}^2) \qquad \mbox{\rm or} \\ 
	|\tilde s_c^p(\bk_m,\bk_n)|^2
	& = & \exp( - R_p Q_{mn})  \qquad \mbox{\rm or} \\ 
	|\tilde s_c^p(\bk_m,\bk_n)|^2
	& = & a_p  (R_p Q_{mn})^{b_p}  \qquad \mbox{\rm etc ...  } .
\eea
	In the above equations, subscripts $_i$ and $_p$ index the 
	parameters belonging to the incoherent or to the partially
	coherent components of the core, and $Q_{mn}$ stands for
	certain experimentally defined relative momentum component
	determined from $\bk_m$ and $\bk_n$. 
	  
	A straightforward counting yields that in the limiting case
	when all momenta
	are equal, the simple formula of
	eq.~(\ref{e:lamnfp_sim}) follows from the shape of the
	$n$-particle Bose-Einstein correlation functions of
	eq.~(\ref{e:fpmix}), as
	$\tilde s^i_c(i,i)  = \tilde s^p_c(i,i) = 1. $

\section{Application to three-particle correlation data}

	As an application of the above formalism, we attempt to
	determine the core fraction $f_c$ and the partially
	coherent fraction $p$ from the strength of the 
	NA44 two - and three- particle correlation functions, $\lambda_{*,2}$
	 and $\lambda_{*,3}$,
	in the CERN SPS S + Pb reactions.
	The two experimentally determined values are 
	$\lambda_{*,2} = 0.44 \pm 0.04 $ and 
	$\lambda_{*,3} = 1.35 \pm 0.12 $ (statistical errors only).~
	\footnote{\samepage Coulomb corrections are large in heavy ion collisions and
	the value of $\lambda_{*,3}$ was determined with the
	help of a newly developed Coulomb 3-particle wave-function
	integration method described in ref.~\cite{alt-3}.}
	Figure 1 illustrates the 2 $\sigma$ contour plots
	in the $(f_c,p_c)$ plane, that is obtained for these
	parameters from the
	experimental values of $\lambda_{*,2}$ and $\lambda_{*,3}$.

\begin{figure}
  \begin{center}
  \epsfig{file=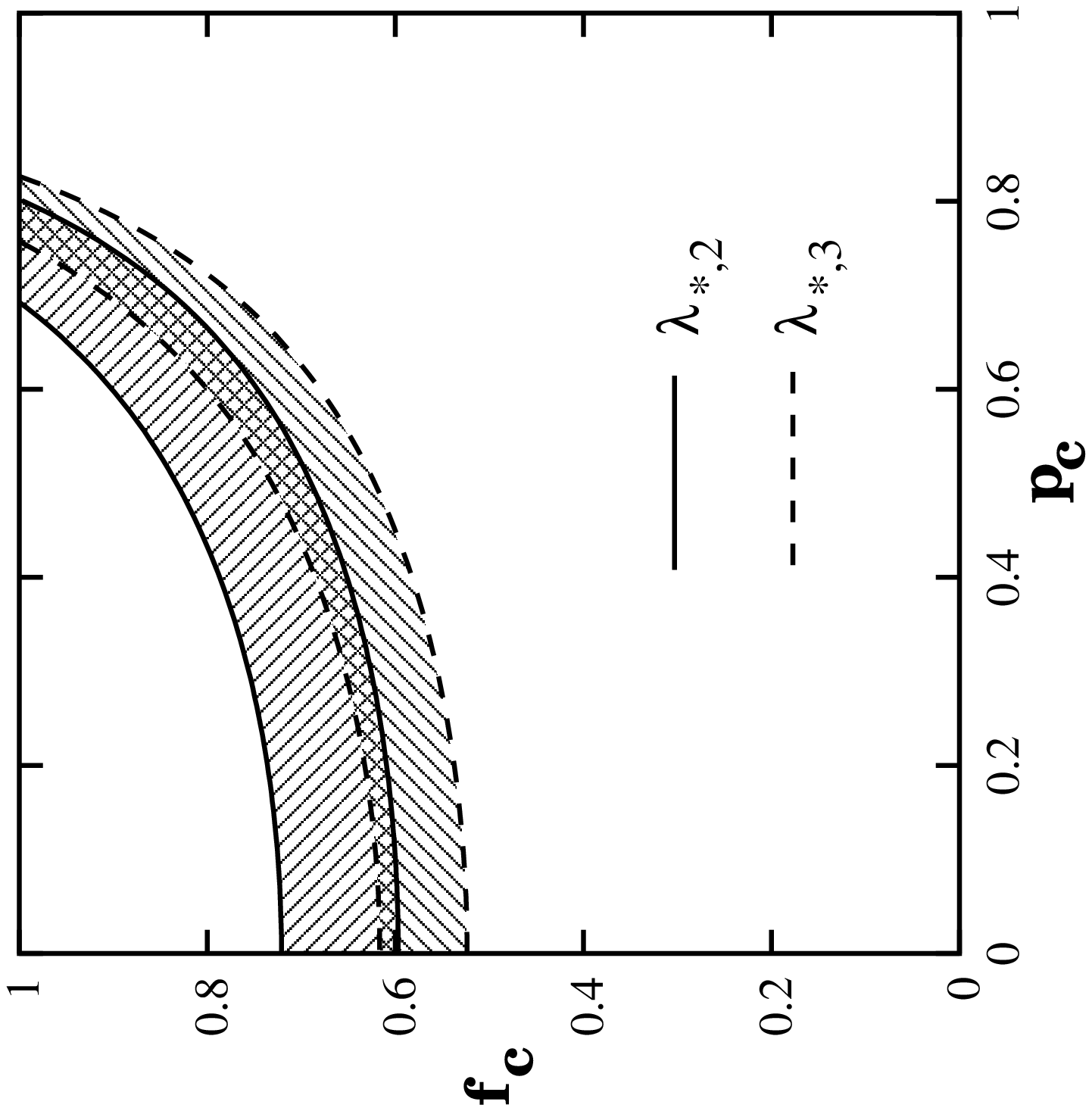,height=10.cm,width=10.cm,angle=270} 
  \end{center}
  \caption{Allowed regions for possible values of the 
  $f_c$ core fraction and the $p_c$ partially coherent fraction
  are evaluated on the two $\sigma$ level from the intercept
  of the second order and the third order correlation
  functions, $\lambda_{*,2}  $ and $\lambda_{*,3}$.
  }
\end{figure}
  The overlap area in
  Fig. 1 shows, that a big range of $(f_c,p_c)$
  values is able to describe simultaneously
  the strength of the two-particle and the three-particle
  correlation functions within two standard deviations 
  of experimental errors. 
	Thus neither the fully incoherent, nor the partially coherent
	source picture can be excluded.

\begin{table}
\begin{center}
\begin{tabular}{|l|l|l|l|l|l|}
\hline
\hline
$f_c$  &  $p$  & $\lambda_{*,2}$ & 
$	\lambda_{*,3}$ & $\lambda_{*,4}$ &$ \lambda_{*,5}$\\
\hline
0.60  &  0.00  &  0.36  &  1.5  &  5.1  &  17.2	\\
0.70  &  0.50  &  0.37  &  1.4  &  4.3  &  11.9	\\
1.00  &  0.75  &  0.44  &  1.6  &  4.3  &  10.5	\\
\hline
\hline
\end{tabular}
\end{center}
\caption{Evaluation of the strength of higher order 
correlation functions, $\lambda_{*,n}  $, for various core fractions and 
partially coherent fractions allowed by NA44 two- and three-particle
correlation data.}
\end{table}
        Now we can predict the intercept of higher order correlations
        to see if they become more sensitive to the presence a partial coherent
        source. In Table 1 we have evaluated the the $\lambda_{*,2},
        \lambda_{*,3},\lambda_{*,4},\lambda_{*,5}$-values for some cases
        in the overlap region. We find that the $\lambda_{*,5}$
	is almost a factor of 2 larger for a completely incoherent source,
	than for a partially coherent source with no halo component,
	although within experimental errors both cases 
	describe $\lambda_{*,2}$ and $\lambda_{*,3}$. 
	This is in agreement with Cramer and Kadija, who have pointed out 
	that for higher values
	of $n$ the difference between a partially coherent source
	and the fully incoherent source 
	will become larger and larger~\cite{cramer-kadija}.

	The results presented here imply that the measurement 
	of higher order correlations, 5-th order,
	is necessary to determine the value of 
	the degree of partial coherence of the source in this reaction.

\section{Summary, conclusions}

	In summary, we have found a simple generalization of the 	
	core-halo model for the case when the core has a partially
	coherent component. The strength of the $n$-particle
	correlation function can be evaluated for arbitrary
	value of $n$ with the help of a simple recurrence formula.

	The shape of the $n$-particle Bose-Einstein
	correlation function was determined in terms of the Fourier-transformed
	emission function of the incoherent and the partially
	coherent component of the source. 
	The graph rules for the calculation of these functions
	are summarized and illustrated graphically in Appendix A. 

	We found that the strengths of the second and the third order
	Bose-Einstein correlation functions in the NA44 $S + Pb$ reaction
	at CERN SPS can be accommodated simultaneously both in
	a fully incoherent core picture $(p_c = 0)$ with a halo fraction of 
	$f_c = 0.6 $ as well as in a partially coherent core picture
	that has no halo component, $p_c = 0.75 $ and $f_c = 1.$ 
	However, the strength of the fourth and fifth order 
	correlation functions is shown to be quite
	different in the two scenarios.
	
\section*{Acknowledgements}
	T. Cs. thanks N. Suzuki and W. Kittel for valuable discussions. 
	T. Cs. and A. Ster would like to express their gratitude
	to B. L\"orstad for kind hospitality at University of Lund
	and to W. Kittel for kind hospitality at University of Nijmegen.
	T. Cs. is grateful to M. Gyulassy for kind hospitality at Columbia
	University.
	
	This research was partially supported by the OTKA grant 
	T026435, by the US - Hungarian Joint Fund MAKA 652/1998
	and by the NWO-OTKA grant N25487. 

\vfill\eject

\vfill
\eject
\section*{Appendix: Graph rules}
	A straight-forward calculation of the higher order 	
	Bose-Einstein correlations for a partially coherent core/halo	
	type of systems is possible with the help of a set of
	graph rules that we determine below. 
	Although the graphs we describe  are similar to those
	of ref.~\cite{suzuki}, the rules are different as we have
	multiplicative factors for each vertex (that carry one momentum
	label each) and for each line (that connect two vertex,
	hence carry two momentum labels each).

	Figures 2 and 3 graphically illustrate the rules of calculations
	of the contributions of the incoherent and coherent core components
	to the $n$-particle correlation function $C_n(\bk_1, ... ,\bk_n)$ 
	for the cases $n$ = 2, 3 and 4, respectively.
	Circles can be either open or full. Each circle carries 
	one label (e.g. $j$) standing for a particle  with 
	momentum $\bk_j$. 
	Full circles represent the incoherent core component by a 
		factor $f_c(j) [1 - p_c(j)]$, 
		whereas open circles correspond to 
		the coherent component of the core, a factor of 
		$f_c(j) p_c(j)$, as 
	defined in eq.~(\ref{e:fck} - \ref{e:fik}) and also shown on Fig. 2.

	For the $n$-particle correlation
	function,  all possible $j$-tuples of particles
	have to be found. 
	Such $j$-tuples can be chosen in 
		${\left( \null^{\dst n}_{\dst  j}\right)}$
		different manner. 
		In such a $j$-tuple,
		either each circle is filled, or
	 	the circle with index $k$ is open and the 
		other $j-1$ circle is filled, which gives
		$j+1$ different possibilities.
	All of the permutations   
	that fully mix either $j = 2$, or 3,	... , or $n$
	different elements have to be taken into account
	for each choice of filling the circles. 
		The number of different fully mixing
		permutations that permute the elements
		$i_1, ... i_j$ is given by $\alpha_j$ and  
		can be determined from the recurrence of eq.~(\ref{e:alp}).

	Lines connect two circles (or vertexes) denoted e.g. by $(i,j)$.
	The lines stand for factors
	that depend both on $\bk_i$ and on $\bk_j$.
	Full lines represent incoherent - incoherent particle pairs,
	and corresponds to a factor of $\tilde s_c^i(i,j)$.
	Dashed lines correspond to  incoherent-coherent pairs,
	and carry a factor of $\tilde s_c^p(i,j)$. 
	The lines are oriented, they point from circle $i$ to
	circle $j$, corresponding to the given permutation, that
	replaces element $j$ by element $i$.
	Dashed lines must start from an open circle and
	point to a full circle. 

\begin{figure}
  \begin{center}
  \epsfig{file=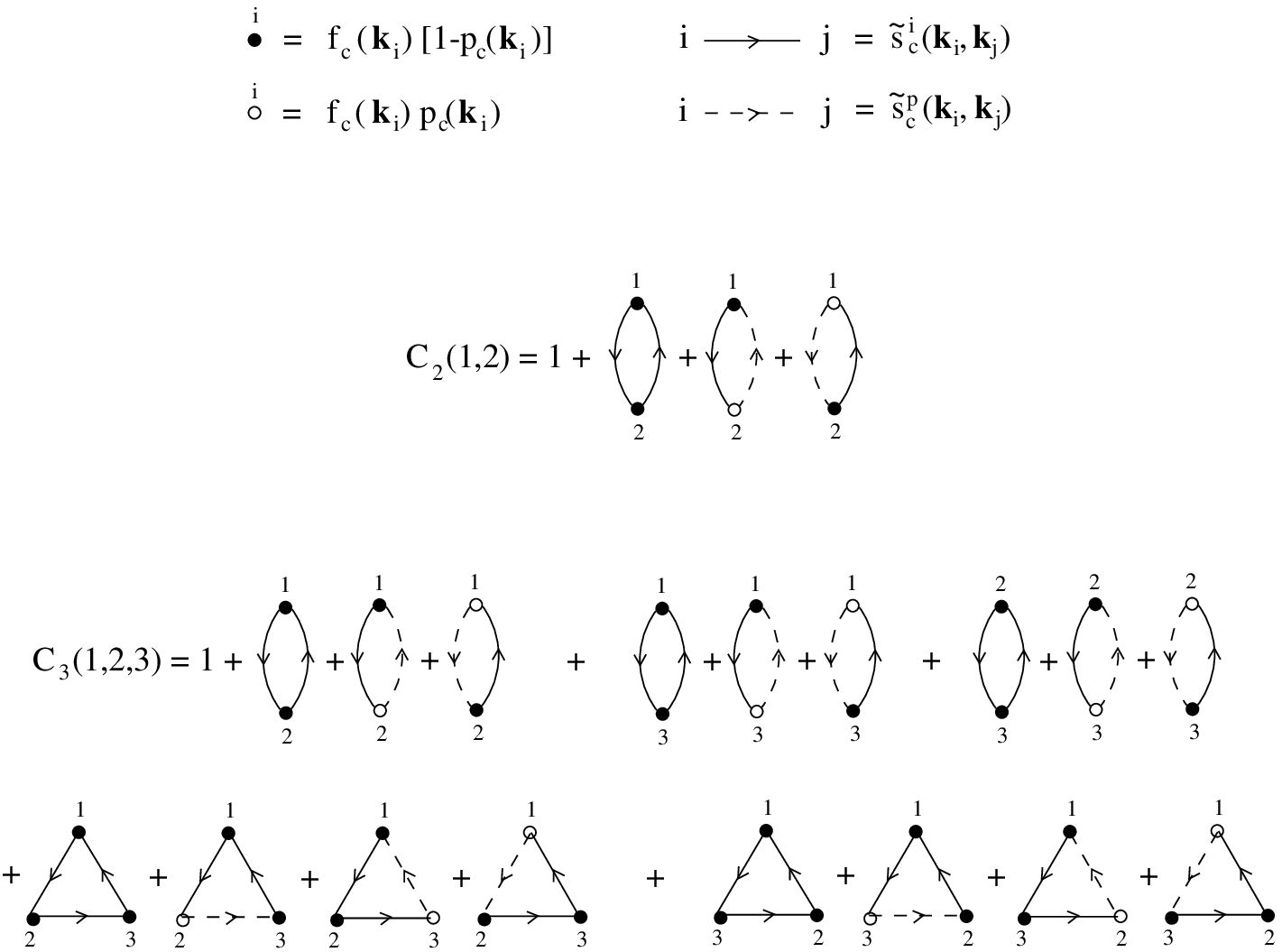,height=8.cm,width=12.cm,angle=0} 
  \end{center}
  \caption{Graphs determining the second and the third order correlation
  function for partially coherent core/halo sources.}
\end{figure}

	All possible graphs must be drawn that are in agreement with
	the above rules. The result corresponds to the
	fully mixing permutations  of all possible $j$-tuples 
	$(j = 2, ... n)$ chosen in all possible manner from
	elements $(1,2, ... , n)$.

	Each graph adds one term to the correlation function, 
	given by the  product of all the factors 
	represented by the cirles and lines of the graph.
	Note that the directions of the arrows matter, as reflected
	by the unequality in eq.~(\ref{e:asym}).
	The correlation function $C(1, ... ,n)$ is
	given by 1 plus the sum of all the graphs.

\begin{figure}
  \begin{center}
  \epsfig{file=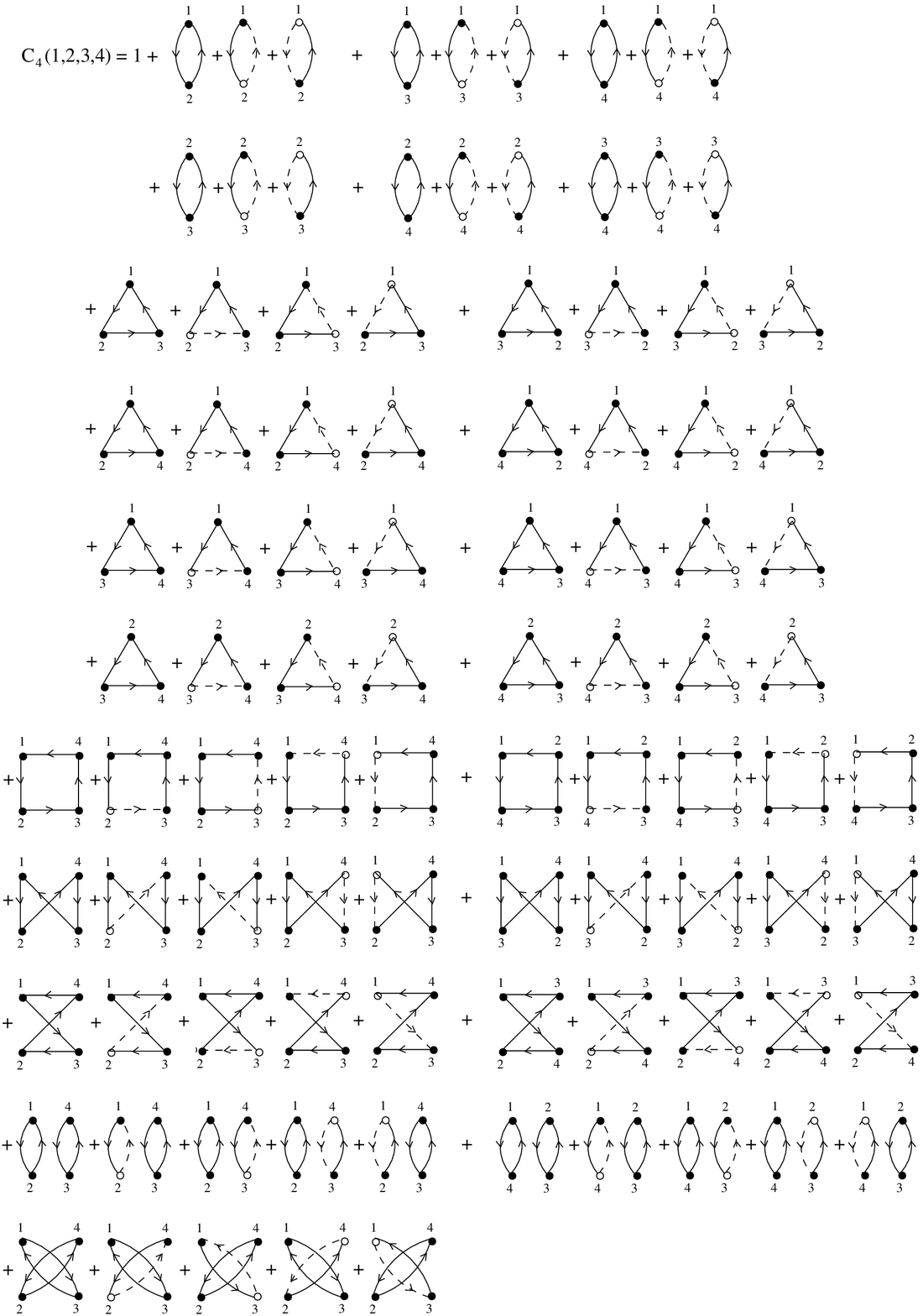,height=15.84cm,width=11.0cm,angle=0} 
  \end{center}
  \caption{Graphs determining the fourth order correlation
  function for partially coherent core/halo sources.}
\end{figure}

	Finally we note, that for the $n$-particle {\it cumulant} correlation
	function, $n$ circles, representing the 
	$n$ particles, should be connected in all possible
	manner corresponding only to the fully mixing permutations
	of  elements $(1, ..., n)$. Disconnected
	graphs do not contribute to the cumulant correlation functions,
	 as they correspond to  permutations,
	 that either do not mix all of the $n$ elements 
	 or can be built up from two or more independent permutations
	 of certain sub-samples of elements $(1, 2, ... , n)$.

\vfill\eject	
\end{document}